# Engineering $Er^{3+}$ placement and emission through chemically-synthesized self-aligned $SiC:O_x$ nanowire photonic crystal structures

Natasha Tabassum,[a,1] Vasileios Nikas,[a,1] Brian Ford,[a] Edward Crawford,[b] and Spyros Gallis [a,*]

[a] College of Nanoscale Sciences and Engineering, State University of New York Polytechnic Institute, Albany, New York 12203, USA

[b] GLOBALFOUNDRIES Corp., East Fishkill, New York 12533, USA

## Abstract

High precision placement and integration of color centers in a silicon-based nanosystem, such as a nanowire (NW) array, exhibiting high integration functionality and high photoluminescence (PL) yield can serve as a critical building block towards the practical realization of devices in the emerging field of quantum technologies. Herein, we report on an innovative synthesis route for realizing ultrathin silicon carbide (SiC) NW arrays doped with and without oxygen ($SiC:O_x$), and also erbium (Er). The arrays of the deterministically positioned NWs are grown in a self-aligned manner through chemical-vapor-deposition (CVD). A key enabler of this synthesis route is that $SiC:O_x$ NW photonic crystal (PC) nanostructures are engineered with tailored geometry in precise locations during nanofabrication. These ultrathin NW PC structures not only facilitate the on-demand positioning of $Er^{3+}$ ions but are pivotal in engineering the emission properties of these color centers. Through a combinational and systematic micro-PL (uPL) and power-dependence PL (PDPL) spectroscopy, PC architecture geometry effects on $Er^{3+}$-related 1538 nm emission, which is the telecommunication wavelength used in optical fibers, were studied. Approximately 60-fold and 30-fold enhancements for, respectively, the room-temperature $Er^{3+}$ PL emission and lifetime in the NW PC sample were observed compared to its thin-film analog. Furthermore, the 1538 nm emission in $SiC:O_x$ NW PC was found to be modulated linearly with the PC lattice periodicity of the structure. The observed characteristics reveal the efficient $Er^{3+}$-emission extraction from the technologically-friendly $SiC:O_x$ NW PC structures.

---

[1] These authors contributed equally to this work

[*] Correspondence and questions should be addressed to S.G. (email: sgalis@sunypoly.edu)

## Introduction

In recent years there has been an enormously high interest in the synthesis, properties, and applications of semiconductor nanowires (NWs). This interest has been complemented by the industry's endless thirst for portable and faster nano-based devices, exhibiting high functionality, with reduced process complexity and energy consumption. The synthesis and simultaneous on-demand-positioning of NWs coupled with the ability to control their orientation and spatial assembly are critical factors towards the production of nano-based devices.[1] The primary limiting challenge commonly faced, especially for feature sizes below 100 nm and bottom-up approaches, is the required deterministic and scalable integration, which involves control over the density, orientation, and spacing of the synthesized nanostructures.

It has been a consensus that semiconductor NWs could incubate the next generation of electronics and photonics,[2,3,4,5] biological and medical technologies,[6,7,8,9] as well as energy technologies.[10,11,12] Following this trend, a great effort has been focused on the development of such nanostructured materials that may be employed ubiquitously in several exciting emerging quantum applications, such as quantum imaging/sensing and quantum photonics.[13,14,15]

Quantum technologies are therefore unavoidably moving towards complicated photonic systems that must be fully integrable and compatible with existing electronic circuits, waveguide architectures and current chip-scale and process technology.[16] Color centers integrated into photonic nanosystems, such as photonic crystal (PC) nanostructures, can experience a redistribution of their spontaneous light emission.[17] By properly engineering the nano-PC, it is possible to control which optical modes are allowed or inhibited to simultaneous suppress and redistribute light emission by the photonic bandgap effect.[18]

Therefore, the realization of practical devices in the emerging field of quantum technologies requires innovative material nanoarchitectures where the optical and quantum properties of color centers can be efficiently engineered and manipulated. Silicon-based wide bandgap nanosystems, such as silicon carbide (SiC),[14,16] with high integration functionality and the deterministic placement of color centers can serve as first critical building blocks towards the implementation of such devices. Additionally, the optically active color centers in these nanosystems must exhibit a radiative transition with high photoluminescence (PL) yield, to enable high fidelity optical readout of the center's quantum state and be potentially suitable for single-photon emission, at room-temperature operation. Realization of such architectures requires the development of advanced synthesis methods and novel nanofabrication integration schemes.[14]

Herein, we present an innovative synthesis route for SiC:O$_x$ NWs that overcomes obstacles usually faced by bottom-up approaches to growing NWs, which typically result in random orientation and size. The arrays of the deterministically-oriented ultrathin SiC NWs were synthesized in a self-aligned manner through a catalyst-free chemical-vapor-deposition (CVD) synthesis route. The resulting NW PC structures are engineered with tailored geometry at precise locations. This integration strategy was extended towards on-demand positioning of photo-stable erbium (Er$^{3+}$) ions and engineering of their emission for applications in optical networks, and quantum science and engineering. Er$^{3+}$ ions were chosen for their intra-4$f$ transition 4$I_{13/2}$ - 4$I_{15/2}$ at the technologically important wavelength of 1538 nm.

## Nanofabrication

### Fabrication of self-aligned ultrathin nanowire arrays

Self-aligned NW arrays grown at predetermined positions were fabricated on silicon substrates via thermal CVD. An abridged fabrication process of the NWs is schematically depicted in Figure 1. First a hydrogen silsesquioxane (HSQ) negative-tone resist layer was spin-coated onto Si. Line patterns were exposed using electron beam lithography (EBL), and the resulting wafer piece was developed in a chemical solution bath, yielding an HSQ ribbon array.

After the development process, 20 nm-thick SiC or SiC:O$_x$ (SiC$_{0.58}$O$_{0.88}$), from now on SiC:O, was conformally deposited onto the HSQ template using CVD (Fig. 1a-ii).[19,20] Following the ultrathin CVD conformal growth, anisotropic reactive ion etching (RIE) was performed to remove the top SiC or SiC:O layer, leaving the sidewall self-aligned layer (NWs) intact and exposing the HSQ template. A wet etch, buffered hydrofluoric acid (BHF), was used to remove the HSQ, yielding ultrathin NW arrays synthesized in a self-aligned manner.

The fabrication process allows for NW array-based architectures that can be surface functionalized.[21] Additionally, this synthesis route allows for NWs with width (W) ≤ 20nm to be fabricated without the use of a lithographic-pattern-transfer technique. The width of the NWs solely depends on the thickness of the conformal sidewall layer. The height (H), lattice periodicity (P$_1$), and sub-lattice periodicity (P$_2$) of the NW array are also shown in Fig. 1c.

## On-demand positioning and integration of $Er^{3+}$ into NW PC structures

Panel b of Figure 1 presents the major process steps employed to engineer the position of $Er^{3+}$ ions in the NWs. A sacrificial thick oxide layer was deposited onto the NW array using CVD,[19] followed by chemical mechanical planarization (CMP) to flatten and thin the oxide layer. Following CMP, a controlled wet etch step using BHF was performed to recess the oxide layer to expose the top of the NWs. A thin (15 nm) conformal CVD-grown oxide layer was then deposited to encapsulate the NWs. The thickness of the oxide layer can be modulated based on the targeted ion implantation depth in the NWs. Accordingly, the remaining thicker oxide layer between the NWs prevents ion implantation into the substrate. Samples were then subjected to a 150 keV beam of erbium ions, targeting ions to be embedded into 30 nm from the top of the NW arrays. Various doses of Er ions – $1\times10^{13}$ cm$^{-2}$ (low), $5\times10^{13}$ cm$^{-2}$ (medium) and $1\times10^{14}$ cm$^{-2}$ (high) were implanted into the NW structures and reference (thin film) samples by ion implantation. A post-implantation argon (Ar) annealing was then carried out at 900°C for 1 hour to activate the Er ions optically.

## Results and discussion

### NW PC architectures modeling

The effect of a two-dimensional (2D) SiC:O NW PC slab on the extraction efficiency of the spontaneous emission (SE) from a dipole oriented along the *y*-axis was explored using finite-difference time-domain (FDTD) calculations. The extraction efficiency is defined as the fraction of the emitted flux through the top and bottom surfaces of the slab along the *z*-direction to the total emitted flux.[22]

FDTD calculations revealed that the extraction efficiency of the reference (thin-film) sample was primarily below 5% at most wavelengths across the simulated range, reaching a maximum of 12.5% around 377 nm (Fig. 2a). The observed behavior was consistent with findings from other studies indicating a high density of guided modes over a broad wavelength range.[22] FDTD calculations similarly performed on SiC:O NW PC nanostructures with three different lattice-periodicity ($P_1$) values revealed two major differences. First, the extraction efficiency was enhanced throughout the whole investigated wavelength range with values up to 38% (Fig. 2a). Specifically, the extraction efficiency for the 1538-nm emission of $Er^{3+}$ was 34% for the 600-nm-$P_1$ NW PC structure, an order of magnitude higher than the reference with ~3.3%. Second, a modulation in the extraction efficiency was observed with the $P_1$ value of the NW PC architectures.

The extraction efficiency enhancement of the emitter's SE in a NW PC structure is expected as a result of the suppression of optical modes in the in-plane directions. This inhibition leads to a redistribution and enhancement of light emission in the *z*-direction.[18,22] This is further justified by the calculations of the spatial distribution of the extracted radiation (~$|E|^2$) at 1538 nm for the reference and the 600-nm-$P_1$ NW PC structure (Fig. 2b). In the case of the reference slab, a significant portion of the radiation was confined within the slab (*x*-direction in this simulation). Conversely, for the 600-nm-$P_1$ NW PC structure, due to the suppression of optical modes in the in-plane direction light emission was inhibited along the *x*-direction, and hence, the radiation appears to have originated primarily from the center of the structure resulting in an enhanced light emission in the vertical direction (Fig. 2b).

**PL properties of SiC:Er and SiC:O:Er NW PC architectures**

The room-temperature $Er^{3+}$-emission ($^4I_{13/2} \rightarrow {^4I_{15/2}}$) properties of the nanoarchitecture (SiC:O NW PC) were investigated by steady-state PL spectroscopy. It is worth highlighting that, with the experimental setup used in this study, the $Er^{3+}$ PL emission at 1538 nm was not detected from a representative reference sample for pumping powers below 30 mW. This power can be defined as the threshold (lowest) power for detecting $Er^{3+}$ PL from the reference. In contrast, $Er^{3+}$ PL from the 600-nm-$P_1$ NW PC architecture was measurable with pumping power as low as 0.7 mW. This behavior suggests that the extraction and collection efficiency of the $Er^{3+}$ emission from the NW PC structure is higher compared to its thin-film analog. At the threshold power, and accounting for the same number of Er ions (see Methods), the $Er^{3+}$ PL collected from the 600-nm-$P_1$ NW PC architecture was found to be approximately 60 times higher than that of its representative reference sample without a PC structure (Fig. 3a). Moreover, at threshold pumping, the PL intensity ratio between the NW PC and reference samples increased monotonically with decreasing Er dose. For example, the PL ratio was approximately 17 and 35 for, respectively, the $10^{14}$ cm$^{-2}$ and $5\times10^{13}$ cm$^{-2}$ doses (inset in Fig. 3a).

To obtain more insight into these compelling behaviors and to further quantify the $Er^{3+}$-emission efficiency in NW PC, as compared to the reference, the $Er^{3+}$ PL was measured as a function of excitation power (PDPL - Fig. 3b). After subtracting the linear background and the detector dark counts, the 1538 nm-emission data were fitted to the following equation,[23] $I(P) = I_{sat}/(1 + P_{sat}/P)$, where *I(P)* is the PL intensity at a given excitation power *P*, $I_{sat}$ is the saturation intensity and $P_{sat}$ is the excitation power required to yield half of $I_{sat}$. Both $I_{sat}$ and $P_{sat}$ were extracted for a representative reference sample without the PC, showing $I_{sat}$ = 192 ± 10 cps and $P_{sat}$ = 113 ± 13 mW. Figure 3b also presents the results from a representative 600-nm-$P_1$ NW PC architecture, revealing $I_{sat}$ = 4244 ±114 cps and $P_{sat}$ = 6 ± 1 mW. These values indicate an approximately 20-fold increase in both pumping efficiency and output flux of $Er^{3+}$ emission in the NW PC architecture. It may be, therefore, suggested that the excitation efficiency of the NW PC, compared to its reference, was enhanced due to improved coupling of the pump laser with the NW PC structure.

Enhanced emission in the direction normal to the PC structure is accompanied by a longer lifetime as a consequence of the reduced optical modes available for recombination leading to a decrease in the SE rate.[18,24] Time-resolved photoluminescence (TRPL) studies of the 1538 nm emission in the NW PC architecture revealed a lifetime of 2.8 ± 0.2 ms obtained from a single exponential fitting (Fig. 4a-i). Conversely, the PL decay of $Er^{3+}$ in the reference was measured ~90 ± 10 μs for the reference (Fig. 4a-ii). The observed lifetime for the NW PC sample was approximately ~30 times longer than that of a reference. At low pumping power, the PL intensity ration between the NW PC and the reference sample is larger than 30x, suggesting a higher concentration of optically active erbium ions in NW PC.[25,26]

In addition to the enhancement effect of the NW PC structure in the $Er^{3+}$ emission lifetime, the integration of color centers in nanostructured materials may offer supplemental benefits such as decreased coupling to non-radiative recombination centers, thus also, leading to an increase of the observed lifetime. In ultrathin SiC:O NWs having well-passivated surfaces, excited $Er^{3+}$ is expected to be exposed to a smaller number of non-radiative recombination sites compared to their bulk analogs.[21,27] Therefore, the increase of the $Er^{3+}$ lifetime –decrease of the PL decay rate- for the NW PC sample is coupled to the above mechanisms, as the PL decay rate is influenced from the complex interplay between the radiating center and its interactions with the host system-structure and surrounding environment (non-radiative channels).

Furthermore, modulation of the $Er^{3+}$ PL intensity in NW PC structures was observed as a function of the $P_1$ lattice periodicity, similar to the trend seen in the FDTD calculations. Although the density of NWs, which can be translated into the density of $Er^{3+}$ ions, is the highest in the PC structure with $P_1$ = 200 nm, the $Er^{3+}$ PL was observed to be the lowest among the studied NW PC structures (Fig. 4b). As presented in Figure 4b, the $Er^{3+}$ PL increased linearly by a factor of about five as $P_1$ increases from 200 to 600 nm. This correlation was also observed for equivalent silicon carbide (SiC) NW PC structures (Fig. 4b).

These findings further revealed the effects of the photonic crystal on the $Er^{3+}$ emission in ultrathin NW PC architectures. A detailed study is vital to further clarify the specific role of height and geometry/periodicity of the NW structures, and also, the relative depth position of the color centers in NWs in defining the emission characteristics of the centers. Follow-up experiments, including detailed TRPL analysis, will be required to shed more light on the underlying processes for the observed PL characteristics of the Er-doped NW PC structures, and, to achieve a better understanding of the role of the PC architecture-geometry.

## Conclusions

In summary, we present a new nanofabrication scheme for the integration of color centers into novel solid-state host nanosystems. These host systems encompass tailorable photonic crystal SiC or SiC:O NW structures, thus enabling the engineering of the SE characteristics of the centers. FDTD calculations revealed that the extraction efficiency could be substantially enhanced by engineeringly controlling the NW PC structure. Appreciable enhancements for both the room-temperature 1538-nm-emission and also the lifetime in the NW PC structures were experimentally observed compared to their thin-film counterparts. Furthermore, the $Er^{3+}$ PL collected from the NW PC structures linearly increased with the $P_1$-lattice-periodicity of the PC structure. This holistic approach may, therefore, incubate an alternative pathway towards advancements in quantum science and engineering that could be benefited by the exact placement and engineering of the emission of color centers, and the large-scale integration of wide band gap hosts.

## Methods

**Synthesis of NW array structures**

Silicon (100) wafers were spin-coated with hydrogen silsesquioxane (HSQ) negative-tone resist at 1000 rpm followed by a soft-bake at 80°C for 4 minutes, yielding an approximately 130 nm thick HSQ layer. The HSQ layer was exposed using a Vistec VB300 electron-beam lithography tool using line patterns created in Layout Editor. Following the exposure, development was performed in tetramethylammonium hydroxide (TMAH) yielding HSQ ribbon arrays. 20 nm of SiC/ SiC:O was conformally deposited onto the HSQ ribbon arrays using a home-built thermal CVD system at 800°C. CVD-742 (1,1,3,3-tetramethyl-1,3-disilacyclobutane) from Starfire Systems was used as the silicon and carbon source. After the SiC/ SiC:O growth, anisotropic fluorine-based (combination of $CHF_3$ and $CF_4$ gases) reactive ion etch (RIE) was performed using a Plasma-Therm Versalock 700 to remove the top SiC/ SiC:O layer, leaving the sidewall layers intact and exposing the HSQ ribbon template. Following the RIE, a wet-etch using buffered hydrofluoric acid (BHF) was used to remove the HSQ, yielding ultrathin NWs synthesized in a self-aligned manner.

**Device modeling**

The FDTD calculations of the extraction efficiencies were carried out using the commercial available Lumerical Solutions software. A single dipole, oriented along the y-direction, was positioned at the center of one NW sourounded by the NWs array. The values for the thin film and NW geometry and index of refraction, used in the simulations, were experimentally determined by SEM and spectroscopic ellipsometry measurements.

**PL characterization**

A home-built micro-PL (µPL) system – composed of an argon laser (Beamlock 2065-7S) coupled to an FLSP920 spectrometer from Edinburgh Instruments – was utilized for PL and power-dependence PL (PDPL) measurements. The PL and PDPL measurements were performed at room-temperature using 488 nm excitation through a Mitutoyo 50x objective lens (NA = 0.55). The $Er^{3+}$ PL intensity of the thin-films was normalized to the effective area -effective number of implanted Er ions- in NW PC structures, as the number of ions is greater for the thin-films in a fixed laser beam-spot area. Therefore, to compare the $Er^{3+}$ PL spectra between the 600-nm-$P_1$ NW

PC and its thin-film analog, the PL intensity in the thin-film was divided with 12, considering the effective area ratio of the thin-film to the NW structure [~1000 nm/ 80 nm (4 NWs)]. Time-resolved photoluminescence (TRPL) studies were conducted in the FLSP920 spectrometer utilizing a multi-channel scaling technique, with a pulsed flash lamp [~3 µs full width at half maximum (FWHM) and 40 Hz repetition rate for NW PC and 80 Hz for reference sample].

## Author contributions

S.G. perceived and designed the experiments; V. N., N.T., and B.F. carried out the fabrication and ion implantation of NW PC architectures; V.N. performed the FDTD calculations; E.C. conducted the structural characterization of the NW PC structures; N.T. performed the PL experiments and analyzed the experimental data. All authors contributed to the discussion of the manuscript; N.T and S.G wrote the main manuscript text.

## Acknowledgements

This work was supported by the Colleges of Nanoscale Science and Engineering of SUNY Polytechnic Institute and The Research Foundation for the State University of New York. The support is gratefully acknowledged.

# Figures

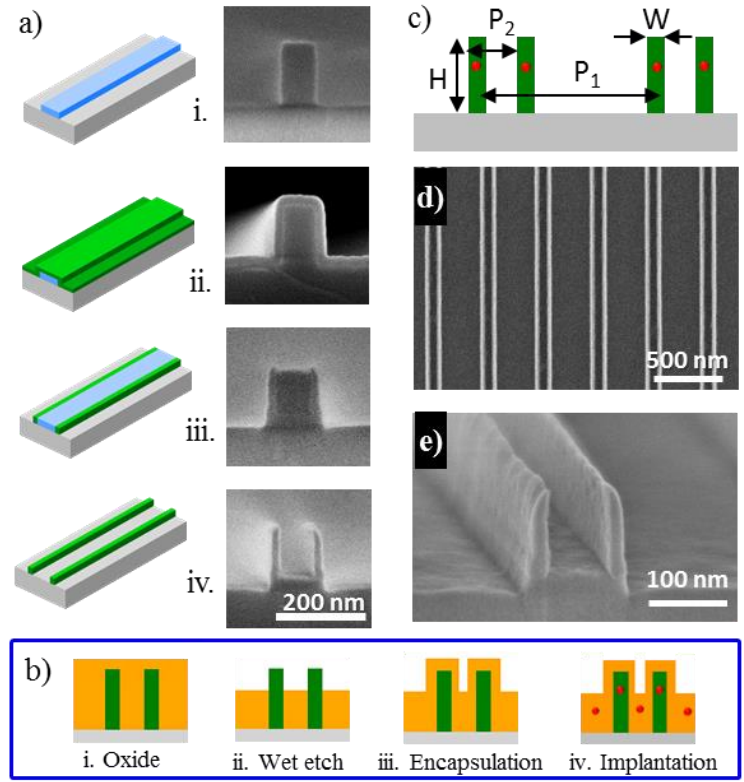

**Figure 1. Integration scheme of ultrathin self-aligned NW PC structures**. (a) Fabrication process-steps for realizing self-aligned 20 nm-width NW PC structures and representative cross-section scanning electron microscopy (SEM) images. i. Spin-coating and EBL of HSQ (blue) to create a ribbon template on a substrate (gray); ii. Self-aligned deposition of conformal SiC or SiC:O ultrathin layer (green); iii. Anisotropic reactive ion etching; iv. Wet-etch removal of HSQ. (b) **On-demand positioning of Er ions.** i. Deposition of sacrificial oxide layer (orange) and CMP; ii. Controlled wet-etch to expose the top of NWs; iii. Deposition of thin oxide (orange) based on the targeted ion implantation depth; iv. Ion (red circles) implantation. (c) Schematic cross-section diagram of the NW PC structures with a PC lattice (pitch - $P_1$) and sub-lattice (pitch - $P_2$); H: Height of the NWs; and W: width of the NWs. (d) Low magnification top-down and (e) High magnification cross-section SEM images of the Er-doped SiC:O NW PC structure (W= 20 nm, $P_1$= 400 nm, $P_2$= 100 nm, H= 130 nm).

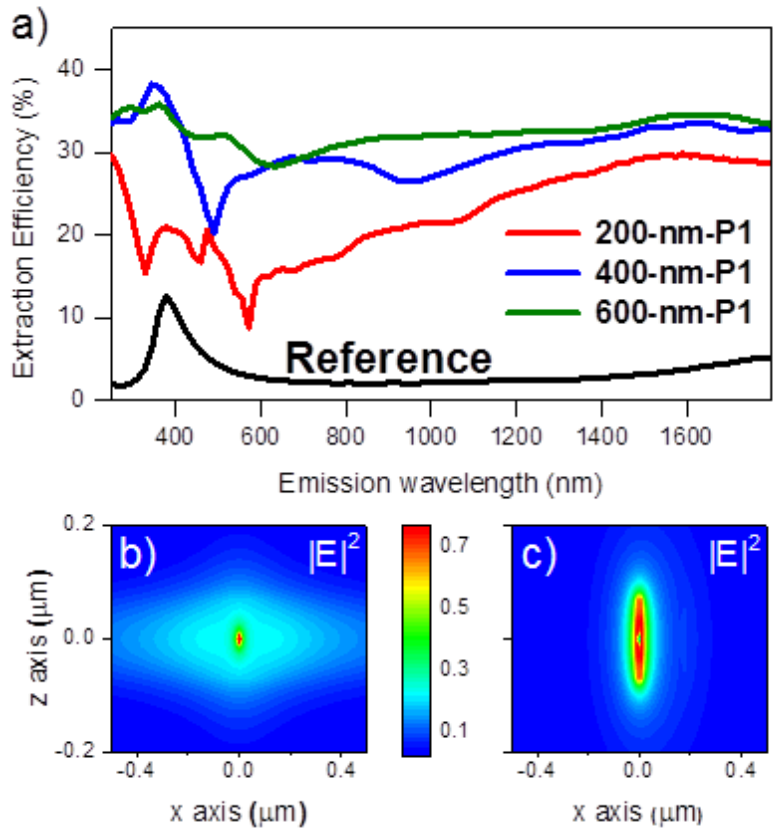

**Figure 2. Device modeling of extraction efficiency for the SiC:O:Er NW PC architectures and reference (thin-film).** (a) A point dipole-source polarized in the *y* direction was inserted inside the SiC:O:Er NW PC structure (W= 20 nm, H= 130 nm) at a 30 nm-depth from the top of the NW; similarly for the reference (thickness= 130 nm). The extraction efficiency for the 600-nm-$P_1$ NW PC structure was calculated to be an order of magnitude higher at 1538 nm emission than its reference analog. **Calculated spatial distribution of the extracted radiation in the *zx* plane** for (b) the reference, and (c) the SiC:O:Er NW PC structure; The radiation was found to be highly confined within the reference (*x*-direction) as opposed to that of the NW PC structure. The electric-field distribution for the 600-nm-$P_1$ NW PC structure is shown to emerge primarily from the center of the structure enhancing radiation in the direction normal to the PC structure.

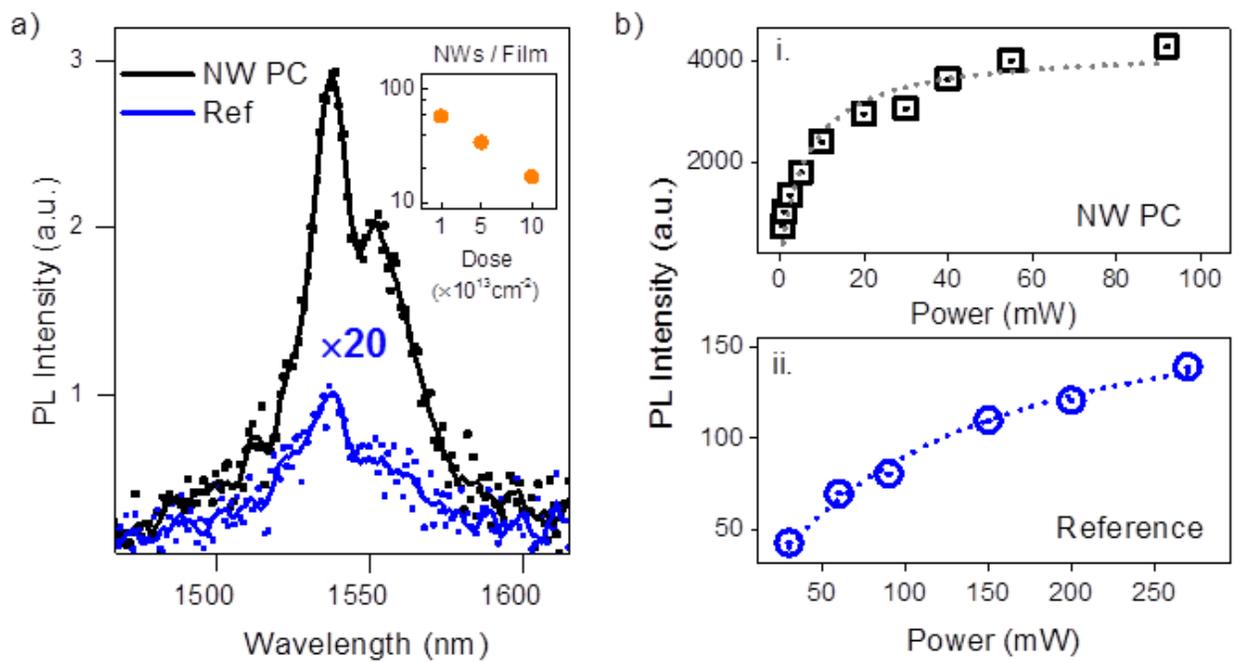

**Figure 3. Effect of the NW PC structure on Er$^{3+}$ emission $^4I_{13/2} \to {}^4I_{15/2}$, ~1538 nm,).** (a) Room-temperature steady-state Er$^{3+}$ PL spectra of SiC:O:Er 600-nm-P$_1$ NW PC and reference samples under 488 nm excitation (Pumping power: 30 mW; Er dose: $10^{13}$ cm$^{-2}$). An approximately 60-fold enhancement for the Er$^{3+}$ PL emission in the NW PC sample was observed compared to its reference analog. The Er$^{3+}$ PL intensity in the reference was normalized to the effective number of implanted Er ions in NW PC (see Methods). Inset: Ratio of Er$^{3+}$ PL intensity between NW PC and reference as a function of Er dose measured at respective threshold power. (b) **Power-dependence PL analysis (PDPL).** Collected Er$^{3+}$ emission as a function of pumping power from i. NW PC and ii. reference. (Er dose: $10^{13}$ cm$^{-2}$).

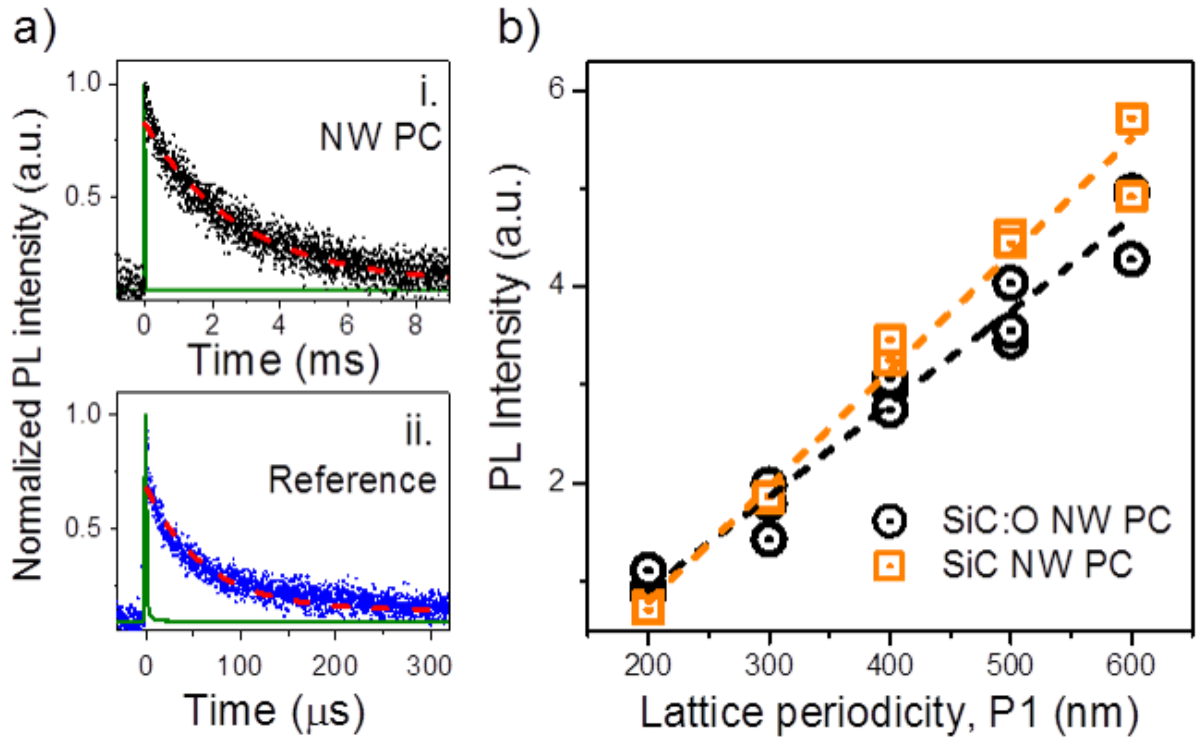

**Figure 4. Dynamics of $Er^{3+}$ PL (~1538 nm).** (a) i. Room-temperature $Er^{3+}$ PL decay from the NW PC structure (black points, Er dose: $10^{14}$ cm$^{-2}$) and instrument response function, IRF (green solid line) under 488 nm pulsed excitation; ii. The PL decay from a representative reference sample (blue points) with corresponding IRF (green solid line). The $Er^{3+}$ PL lifetime was calculated to be ~2.8 ± 0.2 ms for NW PC and ~90 ± 10 μs for the reference sample, as obtained from a single exponential fitting (red dashed lines). **$Er^{3+}$ emission intensity vs NW PC geometry.** (b) Normalized steady-state room-temperature $Er^{3+}$ PL intensity from SiC:Er and SiC:O:Er NW PC structures as a function of $P_1$. Three different NW PC structures were measured for each $P_1$; the dashed lines are the linear fits of the displayed data.